\documentclass{aa}

\usepackage{graphicx}
\usepackage{natbib}
\bibpunct{(}{)}{;}{a}{}{,}
\usepackage{txfonts}

\newcommand{\ms}{${\rm m}\, {\rm s}^{-1}$}
\newcommand{\kms}{${\rm km}\, {\rm s}^{-1}$}
\newcommand{\degree}{$^{\circ}$}

\begin{document}

\title{Retrieval of solar magnetic fields from high-spatial resolution filtergraph data: the Imaging Magnetograph eXperiment (IMaX)}

\author{D.\ Orozco Su\'arez \inst{1,2}, L.R.\ Bellot Rubio \inst{1}, V.\ Mart\'{\i}nez Pillet \inst{3}, J.A.\ Bonet \inst{3}, S.\ Vargas Dom\'{\i}nguez \inst{4}, \\ \and J.C.\ del Toro Iniesta\inst{1}}

\institute{Instituto de Astrof\'{\i}sica de Andaluc\'{\i}a (CSIC), Apdo.\ Correos 3004, 18080 Granada, Spain \and National Astronomical Observatory of Japan, Mitaka, Tokyo 181-8588, Japan \and Instituto de Astrof\'{\i}sica de Canarias, V\'{i}a L\'actea, s/n, 38200 La Laguna (Tenerife), Spain \and UCL - Mullard Space Science Laboratory, Holmbury St Mary, Dorking, Surrey, RH5 6NT, UK\\
 \email{d.orozco@nao.ac.jp; lbellot@iaa.es; vmp@iac.es; jab@iac.es; svd@mssl.ucl.ac.uk; jti@iaa.es} }

\authorrunning{Orozco Su\'arez et al.\/} 
\titlerunning{Retrievals from high-spatial resolution filtergraph data} 

\date{\today}

\abstract {The design of modern instruments does not only imply 
thorough studies of instrumental effects but also a good understanding
of the scientific analysis planned for the data.}  {We investigate the
reliability of Milne-Eddington (ME) inversions of high-resolution
magnetograph measurements such as those to be obtained with the
Imaging Magnetograph eXperiment (IMaX) aboard the {{\sc Sunrise}}
balloon. We also provide arguments to choose 
either \ion{Fe}{i} 525.02 or 525.06~nm as the most suitable line for
IMaX. } {We reproduce an IMaX observation using magnetoconvection
simulations of the quiet Sun and synthesizing the four Stokes profiles
emerging from them.  The profiles are degraded by spatial and spectral
resolution, noise, and limited wavelength sampling, just as real IMaX
measurements. We invert these data and estimate the uncertainties in
the retrieved physical parameters caused by the ME approximation and
the spectral sampling.}  {It is possible to infer the magnetic field
strength, inclination, azimuth, and line-of-sight velocity from
standard IMaX measurements (4 Stokes parameters, 5 wavelength points,
and a signal-to-noise ratio of 1000) applying ME inversions to any of
the \ion{Fe}{i} lines at 525~nm. We also find that telescope
diffraction has important effects on the spectra coming from very high
resolution observations of inhomogeneous atmospheres. Diffration
reduces the amplitude of the polarization signals and changes the
asymmetry of the Stokes profiles. } {The two \ion{Fe}{i} lines at 525
nm meet the scientific requirements of IMaX, but \ion{Fe}{i} 525.02 nm
is to be preferred because it leads to smaller uncertainties in the
retrieved parameters and offers a better detectability of the weakest
(linear) polarization signals prevailing in the quiet Sun.}

\keywords{Magnetic fields -- Sun: photosphere 
-- Instrumentation: high angular resolution}

\maketitle


\section{Introduction} 
\label{sec:intro}

With the advent of new instrumentation, we are witnessing a true
revolution in our understanding of the quiet-Sun magnetism. The
combination of superb spatial resolution and high-precision
polarimetry has unveiled a panorama in which internetwork regions
appear to be covered by very inclined, weak magnetic fields that had
hitherto been elusive to observation. These findings have been
possible thanks to an increase in the spatial resolution of
spectropolarimetric measurements.  Resolutions better than, say,
0\farcs5 were hardly reachable with ground-based polarimeters just
five years ago. Since 2006, however, the \emph{Hinode}
spectropolarimeter
\citep[SP;][]{2001ASPC..236...33L,2007SoPh..243....3K,2008SoPh..249..167T},
is providing data with spatial resolutions of 0\farcs32 on a regular
basis, with a signal-to-noise ratio (SNR) of about 1000. The analysis
of these measurements has shown that the weak internetwork fields
occupy fractions of $\sim 20 - 45$\% of the SP pixel
\citep{2007PASJ...59S.571L, 2008ApJ...672.1237L,
  2007ApJ...670L..61O,2010OrozcoSolenCalma,orozcotesis}. Thus, we are
nearly resolving the magnetic fields that form the internetwork, 
  in the sense that the filling factor is approaching 100\%. But these
  fields are not yet directly visible with current instruments and
their properties have to be determined by means of careful analyses.

Further improvements in the spatial resolution will conceivably lead
to new discoveries. Since the polarization signals scale linearly with
the filling factor, they will become larger if the internetwork is
unevenly permeated by magnetic structures not much smaller than the
current resolution limit. On the contrary, if the characteristic sizes
are very small and the structures homogeneously distributed, then the
observed polarization amplitudes will not increase at higher spatial
resolution because the amount of cancellation of opposite polarity
signals will remain the same\footnote{Traditionally, the cancellation
  of Zeeman signals has been associated with the presence of
  mixed-polarity fields at the spatial resolution of several arcsec
  achieved by Hanle measurements of the quiet Sun \cite[see,
  e.g.,][]{2004Natur.430..326T}.}.  There might be a third
  possibility that the magnetic structures are small but fractally
  distributed \citep{2009ApJ...693.1728P}. In this case, the
  cancellation of opposite polarity signals has been observed to
  depend on the spatial resolution at least down to the 0\farcs3 of
  {\em Hinode}\footnote{This effect was previously shown in
  simulations by \cite{2003ApJ...585..536S}.}. The subject is currently
feeding an intense debate. A distinction between the scenarios and the
determination of their relative contributions to the total magnetic
flux is very important and can effectively be achieved by using larger
telescopes and advanced techniques for high resolution
spectropolarimetry.

In moments of very good seeing, the CRisp Imaging Spectro-Polarimeter
\citep[CRISP;][]{schar06}, installed at the Swedish Solar Telescope on
La Palma (Spain), deliver polarimetric images near the diffraction
limit of its 1-m telescope with the help of adaptive optics and
computationally-expensive image restoration techniques
\citep{2005SoPh..228..191V}. CRISP is a filter-based instrument, which
allows the observation of large fields of view while preserving high
spatial resolution and polarimetric accuracy. Compared to
spectrograph-based polarimeters, filter instruments also have
drawbacks like poorer spectral resolution and wavelength sampling.
Several lines can be observed sequentially, but at the expense of
slower cadences (without degrading the SNR). These advantages and
disadventages imply differences in the results obtained with each
instrument that deserve attention, especially when a new one is
designed and constructed. The way the instrument treats the
polarization signals, the pointing stability, the wavelength samples,
the magnetic sensitivity of the selected spectral line, the
polarization modulation scheme, the polarimetric efficiency, and the
influence of environmental (instrumental plus atmospheric)
polarization are some of the issues to be considered.

The Imaging Magnetograph eXperiment (IMaX) is a dual-beam vector
spectropolarimeter based on a Fabry-P\'erot \'etalon as the spectrum
analyzer and a pair of liquid crystal variable retarders as the
polarization modulator (\citealt{Pillet}; see also 
\citealt{2003SPIE.4843...20J}, \citealt{2004SPIE.5487.1152M}, and
\citealt{2006SPIE.6265E.132A}).  IMaX is one of the post-focus
instruments of the {\sc Sunrise}~1m~telescope, which flew over
  the Artic in a stratospheric balloon from June 8 to 13, 2010
  \citep{Barthol}. In this way, continuous observations of the
Sun were possible avoiding most of the Earth atmosphere.
Table~\ref{cap9:imaxtable} summarizes the basic parameters of the
instrument. IMaX provides capabilities for obtaining polarimetric
images near the diffraction limit of {\sc Sunrise} (0\farcs11 at 525
nm), with a SNR of about $1000$. The instrument has several observing
modes in which the number of wavelengths is varied from 3 through 12
(one of them is always reserved for the continuum), but the
``regular'', so-called vector spectropolarimetric, mode is one that
scans the line at 5 wavelength points in 33~s. Will it be possible to
accurately infer the full vector magnetic field from the regular IMaX
measurements using appropriate inversion techniques?  Answering this
question is the main aim of the present paper.

\begin{table}
\caption{Basic parameters of the {\sc Sunrise}/IMaX system. \label{cap9:imaxtable}}
\begin{tabular}{lr} \hline \hline
{\scshape Telescope aperture} & 1 m \\ 
{\scshape Telescope central obscuration } & 35.2\% \\ 
{\scshape Working wavelength} & 525.02~nm \\ 
{\scshape Spatial resolution}  & $\sim$~0\farcs11\\ 
{\scshape CCD pixel size} &  0$\farcs$055$\times$0$\farcs$055\\ 
{\scshape IMaX passband (FWHM)} &  $\sim$~6~pm \\ 
\hline
\end{tabular} 
\end{table}

To assess the accuracy to which the magnetic field strength,
inclination, azimuth, and line-of-sight (LOS) velocity can be
determined, we here simulate an IMaX observation. We use radiative
magnetohydrodynamic (MHD) models to synthesize the Stokes profiles of
the photospheric \ion{Fe}{i} lines at 525.02 and 525.06~nm. These two
lines, very close in wavelength and hence observable with the same
pre-filter, were considered for the IMaX instrument since the
beginning of the design phase. In fact, part of the results of this
paper provided the rationale for selecting one of the lines instead of
the other. An earlier investigation on magnetic inferences from
filtergrams with limited wavelength sampling was carried out by
\citet{2002SoPh..208..211G}. They used Milne-Eddington (ME) profiles
to simulate the observations. Our paper can be considered as an
extension of their work, since MHD simulations provide a more
realistic description of the solar photosphere.

The simulated Stokes profiles are degraded by telescope diffraction
and detector pixel size to a spatial resolution of 80~km on the solar
surface, and then sampled at five wavelengths (four within the line
plus another in the nearby continuum) to mimic the nominal vector
spectropolarimetric mode of IMaX. From these data we determine the
magnetic field vector and LOS velocity using ME inversions and compare
them with the parameters resulting from the inversion of fully
resolved and critically sampled profiles. To simulate a real
observation, the typical noise of IMaX measurements is added to the
profiles.

The paper is structured as follows: Sect.~\ref{simulandoimax}
describes the spatial degradation of the simulated Stokes profiles and
the effects of the tunable filter on the spectra. In that Section we
also analyze the effects of noise and the drawbacks of selecting only
a few wavelength positions across the
line. Section~\ref{cap9:inversion} deals with the results of ME
inversions. There we estimate the uncertainties of the retrieved physical
parameters. In Sect.~\ref{sec:conclu} we summarize the main
conclusions of this work.


\section{Simulating IMaX observations}
\label{simulandoimax}

\subsection{Spatial degradation}

\begin{figure*}
\centering
\resizebox{0.34\hsize}{!}{\includegraphics{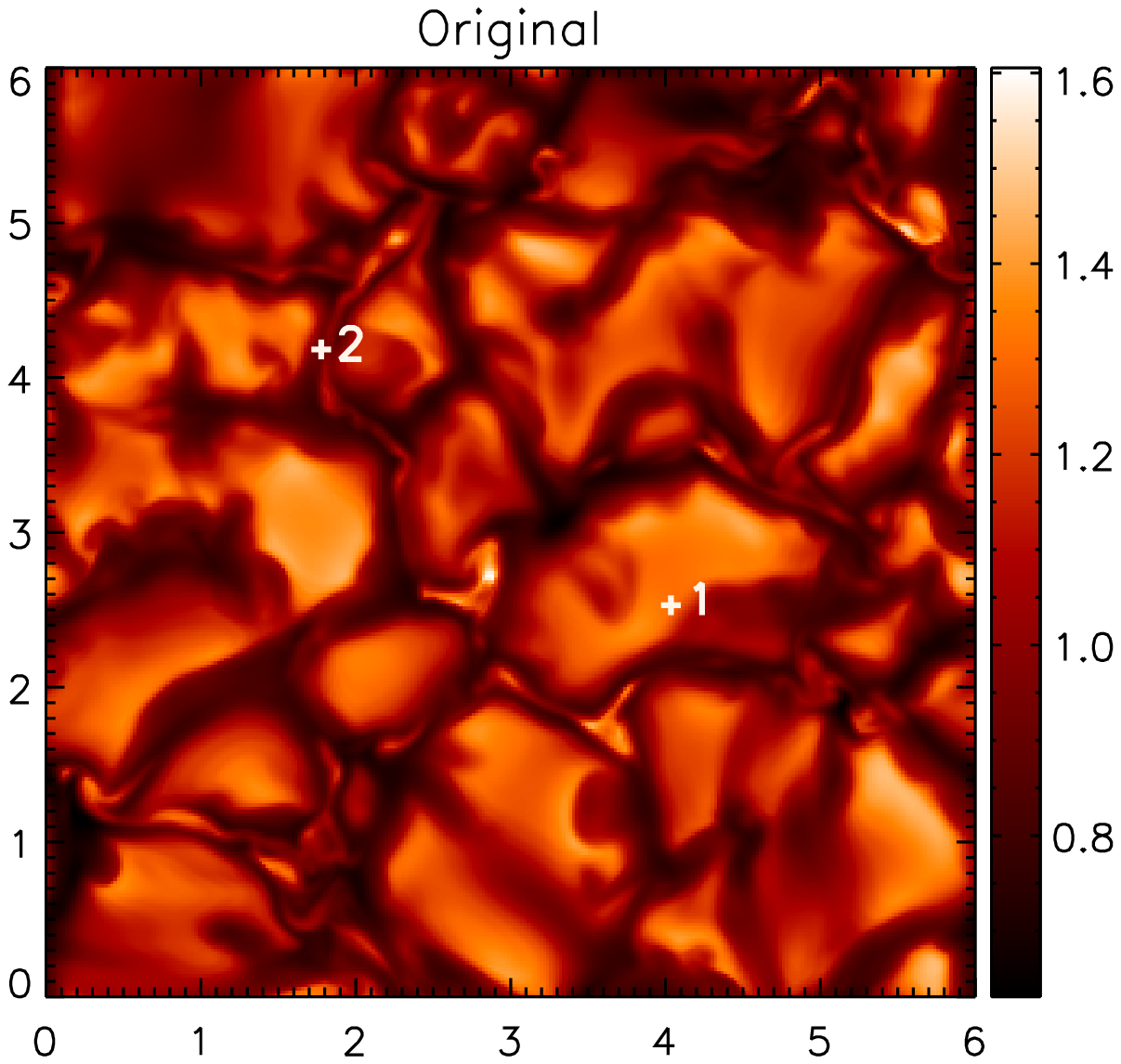}}
\resizebox{0.29\hsize}{!}{\includegraphics{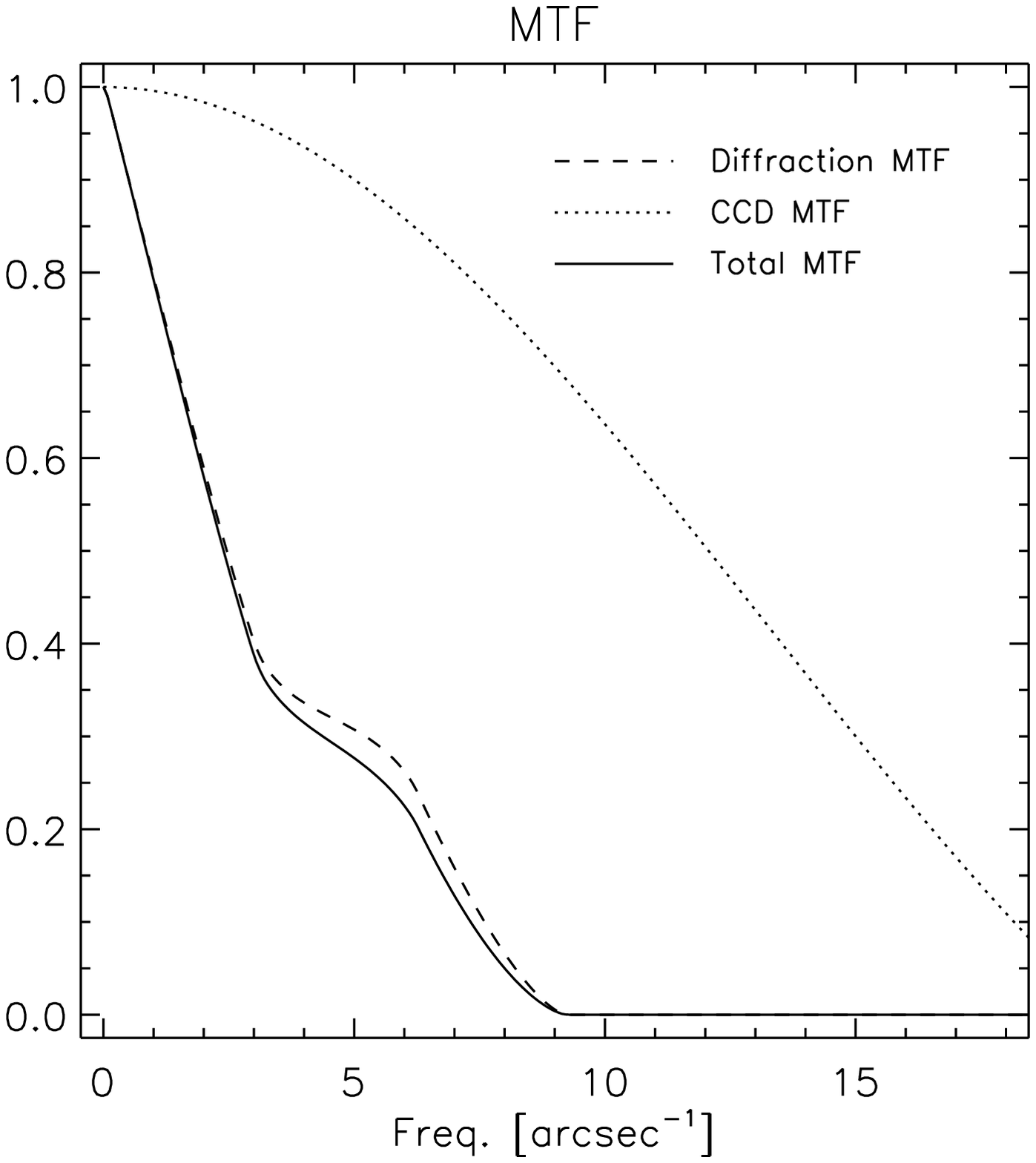}}
\resizebox{0.34\hsize}{!}{\includegraphics{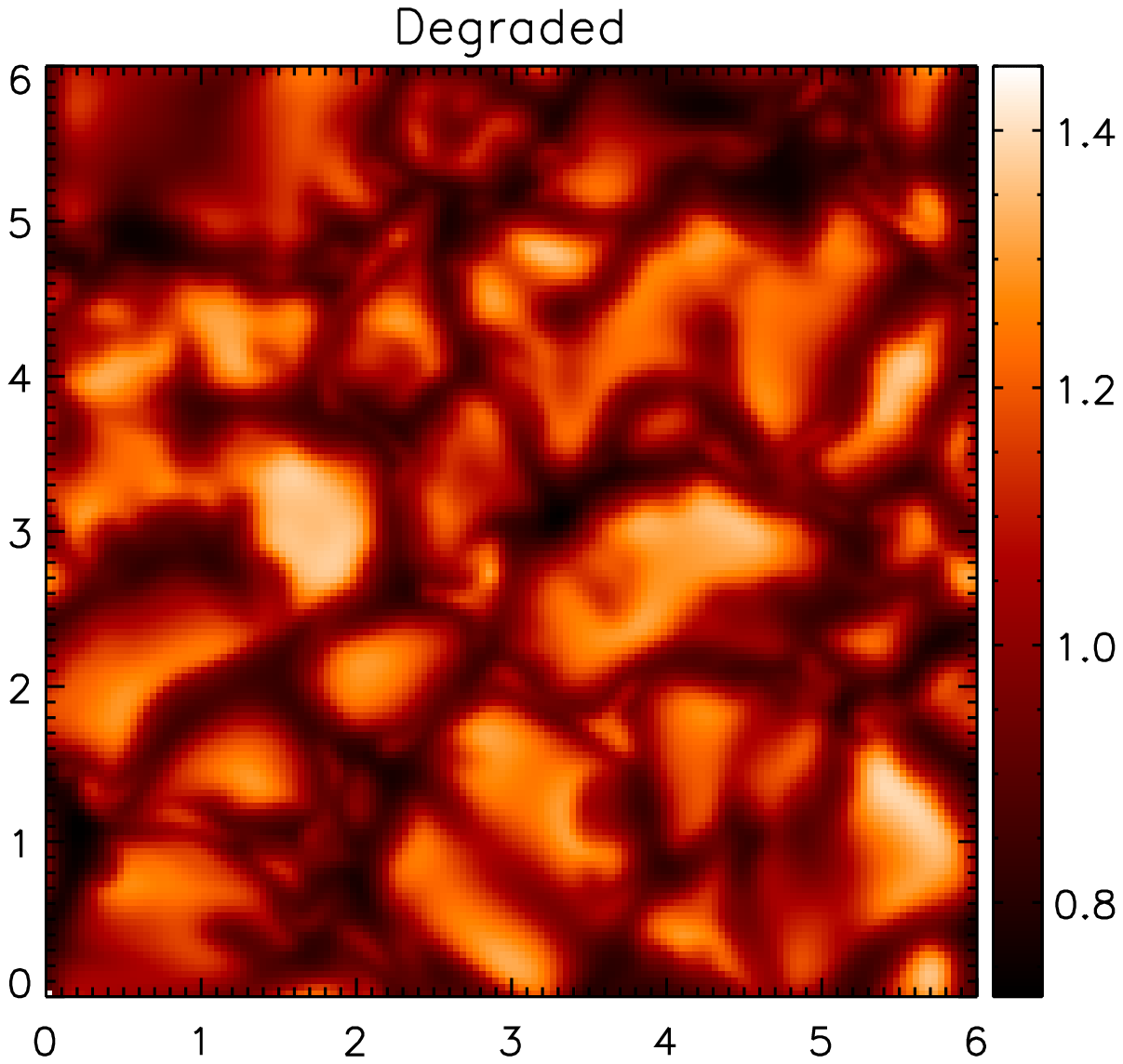}} 
\caption{Continuum intensity map at the original resolution of the MHD models (left) 
and degraded considering telescope diffraction and detector pixel size (right). 
The central panel shows the MTF of the CCD (dotted line), that of the 
diffraction-limited telescope including the central obscuration (dashed line),  
and the total MTF (solid line).} 
\label{fig:continuo} 
\end{figure*}

To simulate an IMaX observation we use model atmospheres from the
radiative MHD calculations of \citet{2005A&A...429..335V}. They
  were initialized with a distribution of mixed polarity fields with
  an average magnetic field strength of $\langle B\rangle\sim200$~G at
  $\log\tau=-1$. The snapshot used here was taken when $\langle
  B\rangle$ had decayed to about $\sim$140~G.  We have chosen this
  particular run because it contains both network- and
  internetwork-like regions. The extent of the computational box is
  $288\times288\times100$ pixels, with a spatial sampling of 20.8~km
  ($0\farcs0287$) in the horizontal direction and 14~km in the
  vertical direction. The bottom of the box is located 800~km below
  $\tau =1$. More details about this run can be found in
  \citet{khomenko}.
  
  Since the work of \citet{2005A&A...429..335V}, simulations of higher
  numerical resolution have become available. They certainly represent
  a significant step forward, but we believe they would not change our
  conclusions, at least for the larger magnetic structures present in
  the network and most of the internetwork -- the ones observed by IMaX
  during its first flight and also by Hinode since its launch in 2006.
  The reason is that the simulations of \citet{2005A&A...429..335V}
  have been confronted with observations thoroughly with excellent
  results. This indicates that they already provide a very good
  description of the solar atmosphere.

The MHD models are used to generate the observations by synthesizing
the Stokes $ I$, $Q$, $U$, and $V$ profiles with the SIR code
\citep{1992ApJ...398..375R}. The spectral synthesis is performed in a
wavelength range that spans 1~nm and includes the \ion{Fe}{i} lines at
525.02 and 525.06~nm. The wavelength step is 1~pm. Next, the
monochromatic images are spatially degraded as the {\sc Sunrise}/IMaX
system does. The modulation transfer function (MTF) of the whole
system is depicted in Fig.~\ref{fig:continuo} and includes the 1 m
aperture of the telescope, the central obscuration of the entrance
pupil (caused by the secondary mirror), and the effects of image
pixelation.  The spatially degraded images are binned by a factor 2,
hence their sampling is almost identical to that of IMaX with a pixel
size of $0\farcs055$. The original continuum contrast of 17.9\%
decreases only by 4\% due to diffraction and CCD pixelation.  After
that we convolve the Stokes profiles with a Gaussian of 6~pm FWHM to
account for the limited spectral resolving power of the Fabry-P\'erot
\'etalon.  Finally, we add noise and select four wavelength samples
across the line plus a wavelength point in the nearby continuum.

The spatial degradation described above represents a rather
pessimistic estimate because IMaX took in-flight calibrations of the
optical aberrations for post-facto restorations.  This will increase
the power at all frequencies up to the cut-off limit. The calibration
procedure \citep{2008Vargasthesis} is based on the phase diversity
(PD) technique for image restoration and wavefront sensing
\citep{1996ApJ...466.1087P}. The instrument has a glass plate that can
be optionally intercalated between the beam splitter and one of the
CCDs to induce a controlled defocus. Therefore, a simultaneous
focus-defocus image pair (PD pair) can be recorded by employing the
two cameras. From this pair, an estimate of the wavefront aberration,
parameterized in the form of a Zernike polynomials expansion, will be
possible during the data processing by means of a PD inversion code.
Deconvolution of the derived point spread function of the system will
be carried out by Wiener filtering the data. Since this will partially
correct the effects of the system, our using the theoretical MTF shown
in Fig.~1 means we are considering a worst-case scenario.

\begin{figure*} 
\begin{center} 
\resizebox{0.8\hsize}{!}{\includegraphics[angle=270]{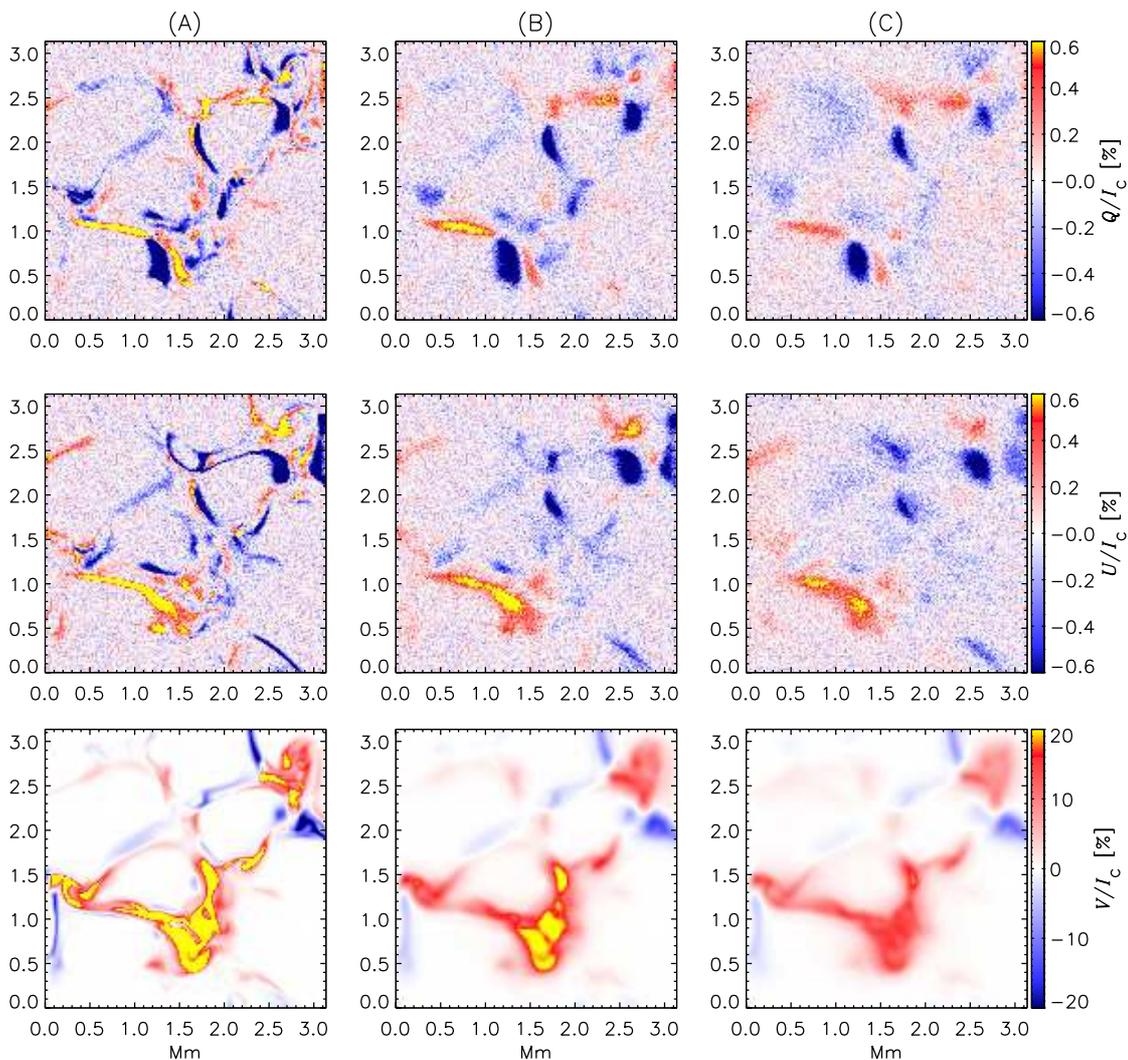}}
\caption{Monochromatic Stokes $Q$ (top), $U$ (middle), and $V$
  (bottom) images at $+7.7$~pm from the central wavelength of the
  \ion{Fe}{i} line at 525.02~nm.  From left to right: the original
  data (A), the spatially degraded data (B), and the spatially and
  spectrally degraded data after taking into account the spectral PSF
  of 6~pm (C). Only a region of $150\times 150$ pixels is displayed.}
\label{fig:QUVimages} 
\end{center} 
\end{figure*}

The two most important effects of telescope diffraction are spatial
blurring and Stokes profile smearing. The first takes place because
(polarized) light spreads from every pixel to its immediate
surroundings. As a consequence, the spatial resolution decreases, the
polarization amplitudes tend to diminish, and the signals are more
vulnerable to photon noise. In addition, the spatial resolution
decreases.  Figure~\ref{fig:QUVimages} displays monochromatic Stokes
$Q$, $U$ and $V$ images at +7.7~pm from the center of the \ion{Fe}{i} 
525.02~nm line where these effects can be observed. To
emphasize details we represent a small area of the simulation
snapshot. (A) stands for the images at original resolution and (B) for
the images spatially degraded by telescope diffraction and CCD
pixelation. We have added noise at the level of
$10^{-3}I_{\mathrm{c}}$ in order to see how it affects the
observations. Since Stokes $Q$ and $U$ have weaker signals than Stokes
$V$, they are significantly more degraded than the latter. In general,
the images do not look strongly altered; the polarization amplitudes
decrease and some structures escape detection, but the morphological
aspect is essentially preserved. 

The second effect comes from the fact that the spatially averaged
Stokes profiles are less asymmetric. As an example, we show in Fig.\
\ref{fig:IVdegradation} the effect of pure spatial degradation
(telescope plus detector) on the polarization profiles. The Stokes $I$
and $V$ signals from two different points (indicated by numbers 1 and
2 in Fig.\ \ref{fig:continuo}) are displayed. Note the strong
modification of the continuum and the shape of the profiles, prior to
any action of the IMaX \'etalon. Similar modifications can be expected
from any instrument, but apparently they have not been described in
the literature. MHD simulations like the ones we are using here are
the best means to date to bring this effect to the attention of the
community: details of the profiles are lost because of the finite
spatial resolving power. 

\subsection{Spectral degradation}

IMaX uses a single Fabry-P\'erot \'etalon in double pass for the
spectral analysis. The FWHM of the IMaX spectral point spread function
(PSF) is $\sim$6~pm. For simplicity, we approximate the shape of the
PSF by a Gaussian function. The effect of the IMaX \'etalon on the
polarization images is shown in column (C) of Fig.\ \ref{fig:QUVimages}. 
Naturally, the amplitudes diminish and the weakest signals disappear 
below the noise.

As is well known, the spectral smearing induced by the \'etalon
smooths out the profile shapes and the Stokes spectra become more
symmetric. Therefore, some of the weakest signals, typically linear
polarization signals, cannot be detected after the combined action of
the telescope, the \'etalon, and the detector. But the problem is not
only on detection. These effects also reduce the information content
of the detected profiles. The sensitivity to gas velocities, for
example, decreases because the \'etalon broadens the profiles and
makes them shallower
\citep{2005A&A...439..687C}. The sensitivity to other atmospheric
parameters also decreases with a wider spectral PSF because response
functions also get broadened by it \citep{2007A&A...462.1137O}.
 

\subsection{The effect of noise}

Both telescope diffraction (plus CCD pixelation) and spectral analysis 
reduce the amplitude of the polarization signals. Consequently, they 
are more affected by the noise. In this section we quantify the 
influence of photon noise on the measurements. 

Figure~\ref{fig:QUnoise} shows the percentage of pixels whose Stokes
$Q$ or $U$ amplitudes exceed specified noise levels. The \ion{Fe}{i}
lines at 525.06 and 525.02~nm are represented in red and black,
respectively. Obviously, the amount of detectable linear polarization
signal decreases quite rapidly with noise. Noise levels of $3\times
10^{-3} I_{\mathrm{c}}$ and $10^{-3} I_{\mathrm{c}}$, the latter
corresponding to the IMaX case, are marked with vertical lines. The
dotted lines stand for the spatially degraded Stokes profiles and the
solid lines for the spatially and spectrally degraded profies. 80\%
and 55\% of the pixels show linear polarization signals in \ion{Fe}{i}
525.02~nm and 525.06~nm that exceed the noise level before the light
enters the instrument. These values change to 50\% and 20\% after the
spectral analysis. Interestingly, the final detectability of linear
polarization in the 525.02~nm line is almost the same as that of
525.06~nm {\em before} the spectral analysis.  Thus, \ion{Fe}{i}
525.02~nm presents clear advantages for detecting weak linear
polarization signals in the quiet Sun.

\begin{figure} 
\centering \resizebox{\hsize}{!}{\includegraphics{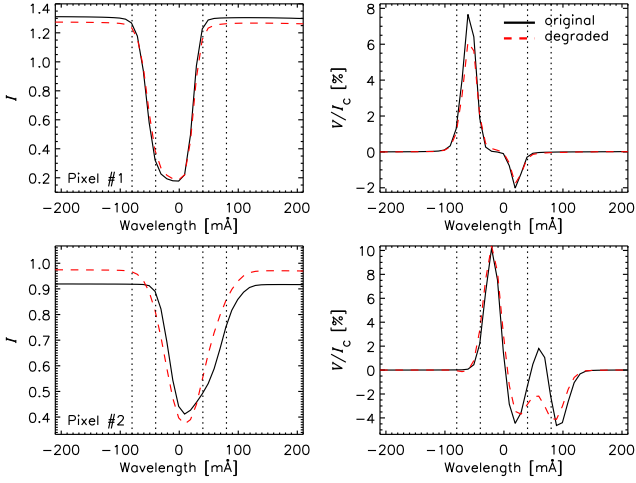}} 

\caption{Stokes $I$ and $V$ profiles of \ion{Fe}{i} 525.02~nm
  corresponding to pixels \#1 (top) and 2 (bottom) in Fig.\
  \ref{fig:continuo}. The solid and dashed lines represent the
  original simulations and the maps degraded by telescope diffraction
  and CCD pixelation, respectively. The vertical lines indicate
  wavelength samples at $[\pm$8,$\pm4]$~pm from the central wavelength
  of the line. To make a pixel-to-pixel comparison, we have taken the
  Stokes profiles from the degraded data before binning the images to
  match the IMaX CCD pixel size. Therefore, dissimilarities are
  enhanced.}
\label{fig:IVdegradation}
\end{figure}

\begin{figure} 
\centering  \resizebox{\hsize}{!}{\includegraphics{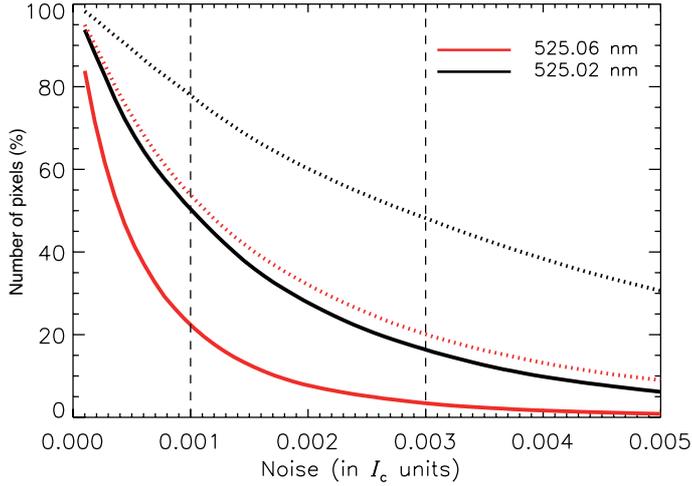}}
\caption{Variation in the percentage of pixels where either Stokes $Q$ 
or $U$ exceed a given noise level. The dotted and solid lines 
correspond to data before and after convolution with the spectral PSF. The vertical lines 
represent noise levels of $10^{-3}$ and $3 \times 10^{-3} I_\mathrm{c}$.}
\label{fig:QUnoise} 
\end{figure}

\begin{figure}
\centering 
\resizebox{0.85\hsize}{!}{\includegraphics{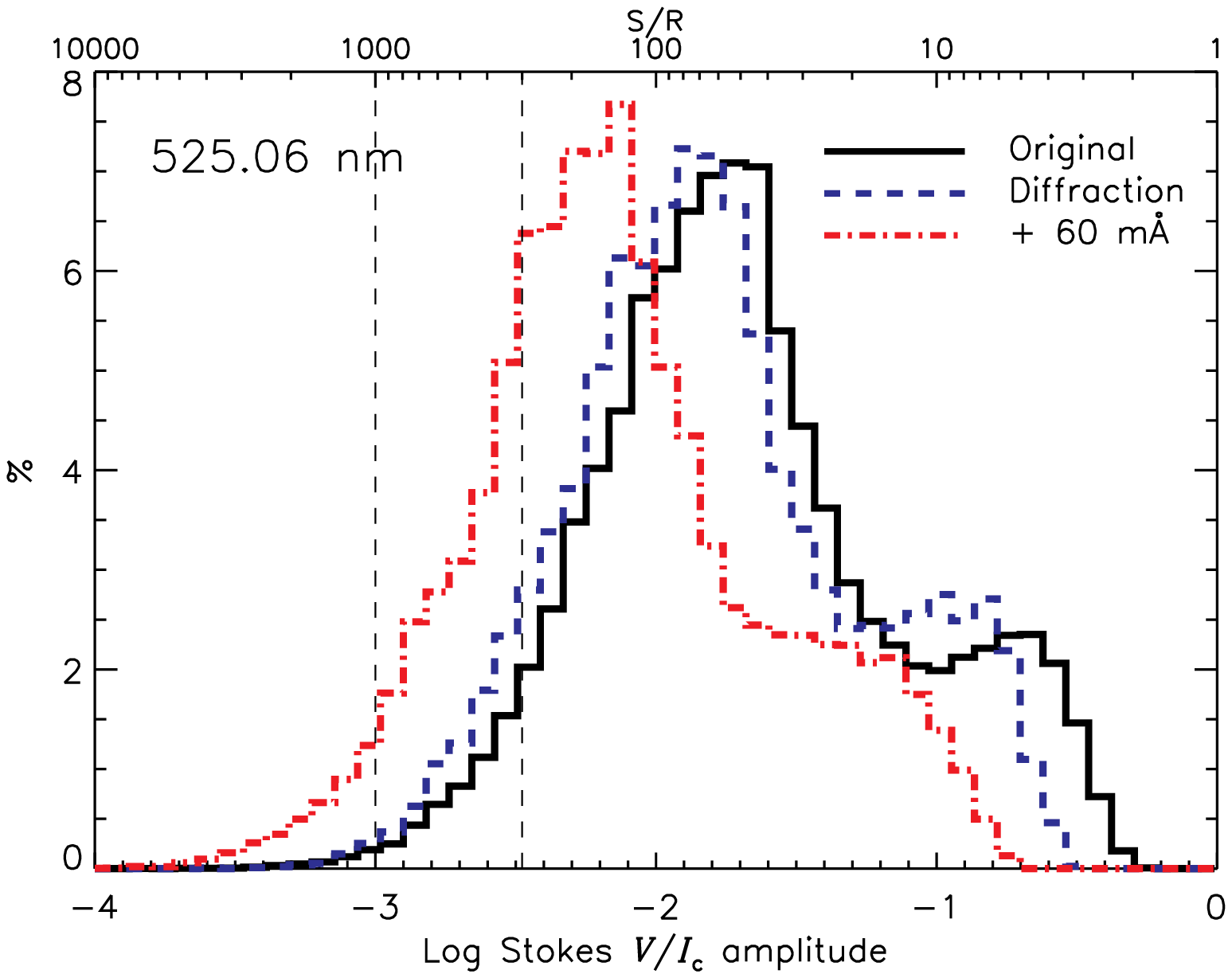}}
\resizebox{0.85\hsize}{!}{\includegraphics{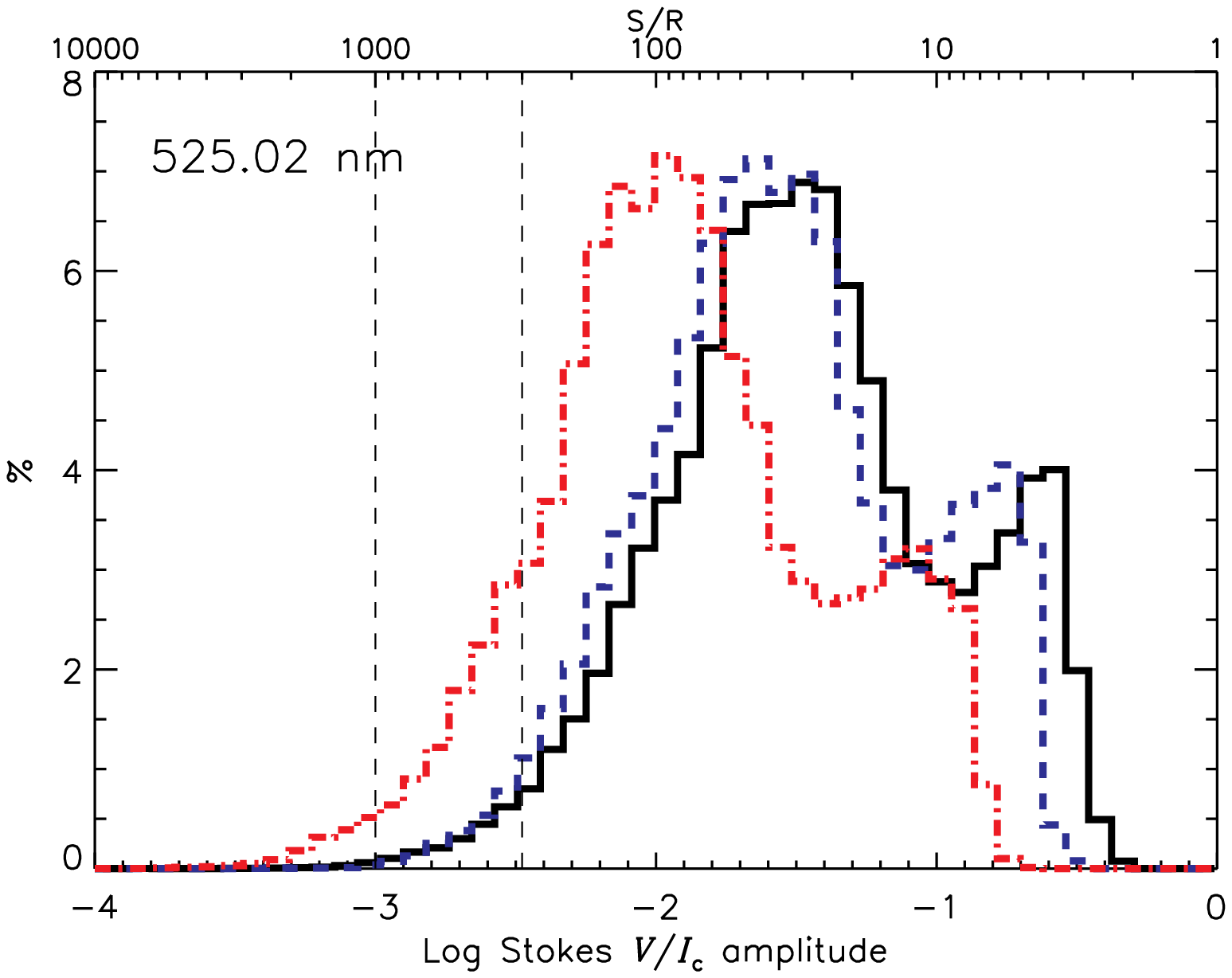}} 
\caption{Distribution of the Stokes $V/I_\mathrm{c}$ amplitudes in the 
original, spatially and spectrally degraded data. The vertical lines 
represent noise levels of $10^{-3}$ and $3 \times 10^{-3} I_\mathrm{c}$.} \label{fig:Vnoise} 
\end{figure}

Figure \ref{fig:Vnoise} shows histograms of the Stokes $V$ amplitudes
in the original maps (solid lines), the spatially degraded maps
(dashed lines), and the spatially plus spectrally degraded maps  
(dash-dotted lines), for \ion{Fe}{i} 525.06 and 525.02~nm (top and
bottom, respectively). The histograms show an asymmetric distribution
of amplitudes with two peaks corresponding to network and internetwork
points, the latter being responsible for the smaller $V$ amplitudes.
The histograms for the 525.02~nm line are shifted to stronger signals
because of the enhanced sensitivity of this line to magnetic fields.
Again, specific noise levels are indicated by the vertical lines. In
the original data, the percentage of pixels with Stokes $V$ amplitudes
below $10^{-3}\, I_{\rm c}$ is 0.06\% and 0.5\% for the 525.02 and
525.06~nm lines, respectively. These percentages change to 1.6\% and
to 4.3\% when the profiles are degraded spatially and
spectrally\footnote{Of course, if ${\rm SNR} = 300$, the Stokes $V$
  detectability decreases: the percentages reach 2\% and 5.3\% in the
  original data, and 11.4\% and 23.2\% after degradation.}. Therefore,
most of the $V$ signals would be observable with our instrument if we
use the 525.02 nm line and the SNR is 1000. Thus, \ion{Fe}{i}
525.02~nm reveals itself as the better choice for detecting both weak
linear and circular polarization signals.

Further degradation is expected due to the limited wavelength
sampling. As mentioned in the introduction, the regular vector
spectropolarimetric mode of IMaX measures the polarization state in 4
wavelength within the line plus another in the continuum. These
wavelengths have been marked with vertical lines in Fig.\
\ref{fig:IVdegradation}. The choice of a limited wavelength sampling 
is not arbitrary: on the one hand, scanning the line at just a few
points reduces the scan time, improving the temporal integrity of the
data and their immunity to variations in the conditions of the dynamic
Sun or changes in the solar scene; on the other hand, a better
sampling was not needed because full line profiles were going to be
obtained with the slit spectropolarimeter originally included in the
scientific payload of {\sc Sunrise}. Here we are concerned with the regular
IMaX spectropolarimetric mode and therefore we have to acknowledge a
further decrease in the fraction of polarization signals detectable
above the noise level.


\section{Inversion of the Stokes profiles}
\label{cap9:inversion}

In this section we address two questions. First, we want to know which
line is less affected by the constraints imposed by the instrument
(i.e., spectral degradation, limited wavelength sampling, photon
noise). The answer to this question is important to choose the most
suitable line for IMaX. Second, we want to determine the uncertainties
that can be expected from the analysis of the measurements taken
in the selected line.

To answer the first question, we need to evaluate the influence of
instrumental effects on the ME parameters resulting from the inversion
of the two spectral lines. This requires a {\em differential}
analysis. In principle, since the original model atmospheres are at
our disposal, it would be straightforward to compare the inversion
results for each pixel and each spectral line as observed by IMaX with
the corresponding MHD models. However, such an {\em absolute} analysis
would not inform us about the relative merits of the two lines. The
reason is that they are formed in slightly different layers of the
atmosphere, and a direct comparison of the ME parameters derived from
them with the MHD parameters at a fixed height would yield differences
that do not reflect the capabilities of the lines, but rather their
different formation heights.

The best way to solve this problem is to set up a reference for
each of the lines. Here, the reference is taken to be the
result of a ME inversion of the spatially degraded Stokes profiles
without noise, sampled at 61 wavelength points separated by 1~pm.
They represent the idealized case in which the instrument would
measure fully resolved Stokes profiles with excellent (infinite)
signal-to-noise ratios. Note that the reference includes the
effects of telescope diffraction, whose influence on the polarization
spectra has already been analyzed in Sect.~\ref{simulandoimax}.

\begin{figure*}[t] 
\centering 
\resizebox{0.7\hsize}{!}{\includegraphics{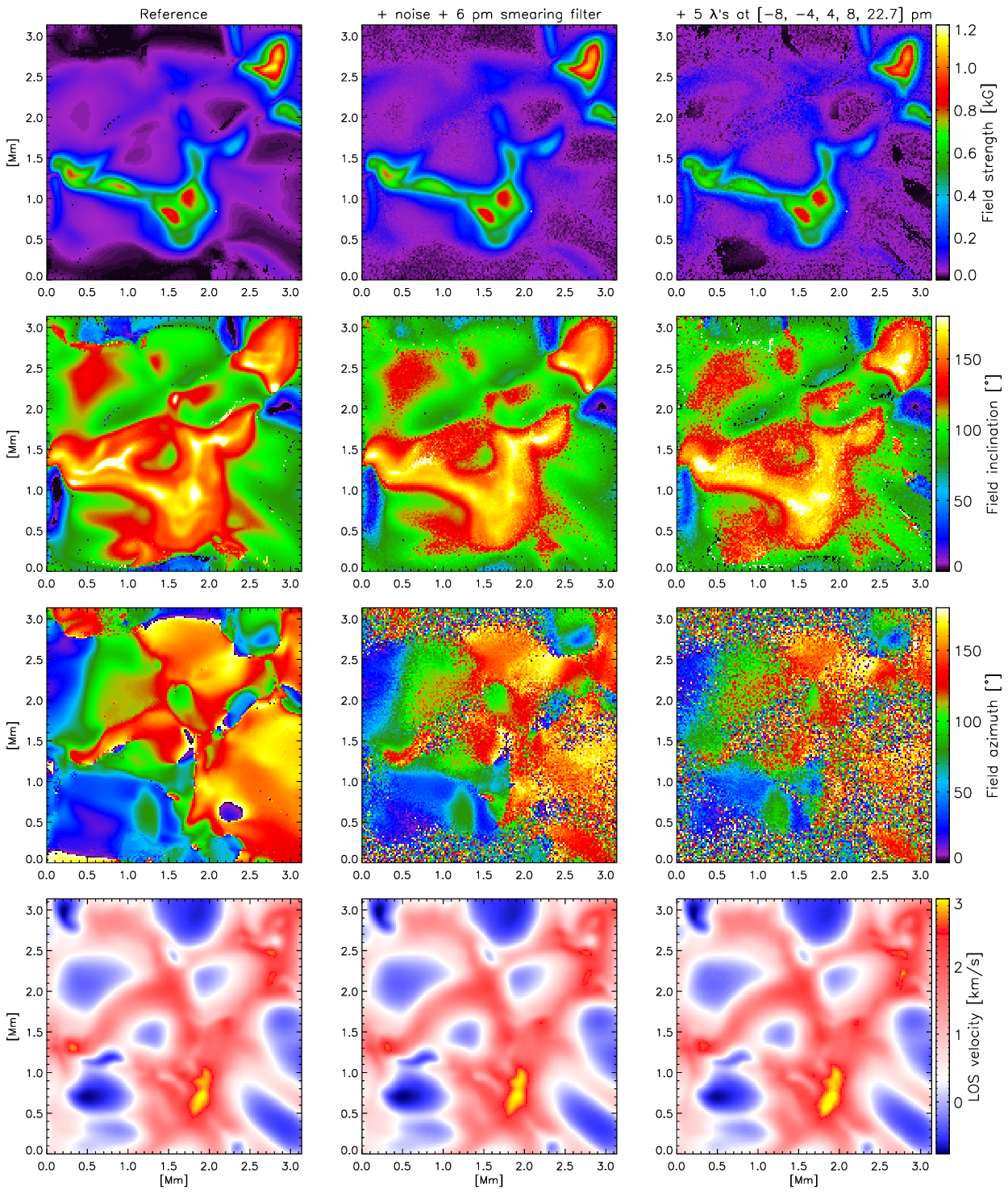}}
  \caption{ME inversions of the \ion{Fe}{i} 525.02 nm line synthesized
  with the help of MHD models. From top to bottom: magnetic field
  strength, inclination and azimuth, and LOS velocity. The left panels
  represent the results from the data degraded by telescope
  diffraction and CCD pixelation. The middle panels show the maps
  retrieved from the degraded profiles after convolution with a 6~pm
  wide filter and addition of noise. The right panels display the
  results obtained from the spatially and spectrally degraded
  observations with noise, sampled at five wavelengths points. Except
  for the reference, ${\rm SNR}=1000$ in all maps.}
  \label{invfull:cap9} \end{figure*}

The {\em differential} analysis is performed by inverting the
simulated IMaX measurements in the two lines and comparing the
inferred ME parameters with the corresponding reference. The results,
presented in Section~\ref{sect3.1}, led to the selection of
\ion{Fe}{i} 525.50~nm as the IMaX line.

Answering the second question requires a comparison of the original
MHD models and the ME parameters derived from the \ion{Fe}{i}
520.52~nm measurements. This is done in Sect.~\ref{sect3.2} below.

All the inversions are carried out assuming a simple one-component
model atmosphere and no stray-light contamination. The magnetic field
vector is initialized with $B=200$ G, $\gamma=20$\degree, and
$\chi=20$\degree. The inversion code used in the tests is MILOS
\citep{2007A&A...462.1137O}.


\subsection{Impact of instrumental effects}
\label{sect3.1}

To determine the deviations induced by the instrument with respect to the reference, we invert the
simulated IMaX data, i.e., the spatially degraded Stokes profiles,
smeared by a spectral PSF of 6~pm FWHM, sampled at 5 wavelength points
(one of which is continuum), and with noise added at the level of
$10^{-3}\, I_{\rm c}$.

The inversion returns the three components of the vector magnetic
field and the LOS velocity. The maps corresponding to the \ion{Fe}{i}
525.02 nm line are displayed in Fig.~\ref{invfull:cap9}. The left
column shows the reference.  The middle column displays the
results of inverting the profiles convolved with a PSF of 6~pm FWHM
and with noise, but still sampled at 61 wavelengths. The right column
gives the results obtained from the simulated IMaX measurements. Only
a region corresponding to $150\times 150$ pixels is displayed.

A comparison of the maps demonstrates that the field strength and
the LOS velocity are well recovered. Only small deviations from the
reference parameters can be seen, although the deviations increase
as the field weakens (especially below 100~G). The inclination and
azimuth maps show the largest deviations because the spectral
smearing, the noise, and the limited sampling reduce the Stokes $Q$
and $U$ amplitudes. The noisier areas in the maps derived from the
simulated IMaX data correspond to regions where the linear
polarization signals are buried in the noise and the field is very
weak.

The overall conclusion drawn from these tests is that ME inversions of
IMaX measurements in the \ion{Fe}{i} 525.02 nm line are reliable. The
accuracy can be expected to be higher for the field strength and the
LOS velocity, but also the magnetic inclination and azimuth, although
noisier, will be recovered reasonably well.

\begin{figure} \centering
\resizebox{\hsize}{!}{\includegraphics{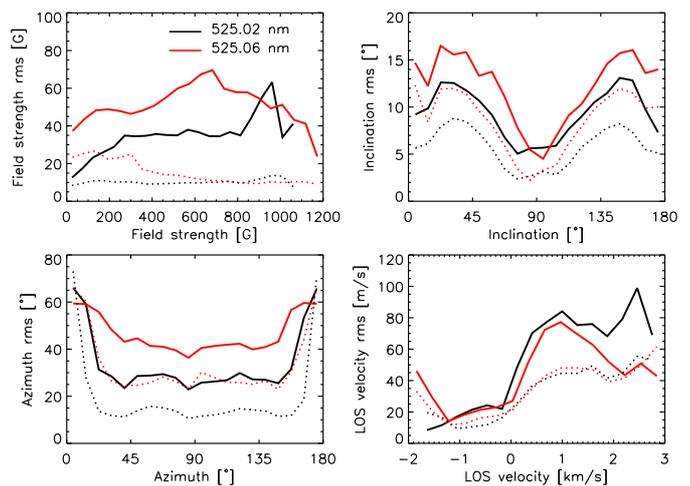}}
\caption{Rms differences between the inferred ME parameters and the
  reference.  The black and red lines stand for \ion{Fe}{i}
  525.02~nm and 525.06~nm, respectively. The dotted lines correspond
  to the inversion of the degraded profiles but with full spectral
  sampling.  The solid lines show the results for the degraded data
  sampled at the five wavelengths of the IMaX vector
  spectropolarimetric mode.}
\label{invfull55:cap9} \end{figure}  

To support this idea more quantitatively, Fig.~\ref{invfull55:cap9}
displays the rms deviations between the ME parameters inferred from
each line and the corresponding reference. The rms values have
been calculated by dividing the x-axis in bins of size 50~G, 9\degree,
and 300~\ms. The black and red lines refer to \ion{Fe}{i} 525.02 and
525.06~nm, respectively. The solid lines indicate the results obtained
from the spatially and spectrally degraded observations sampled at the
five wavelength positions of IMaX (the corresponding maps are the ones
in the right column of Fig.~\ref{invfull:cap9}). The rms values refer
only to pixels in which the Stokes $Q$, $U$ or $V$ signals of
\ion{Fe}{i} 525.06~nm exceed three times the noise level, except for
the LOS velocity where we included all the pixels. Clearly, the
525.02~nm line yields smaller deviations; typical values are 35~G
for the field strength, 10\degree\/ and 35\degree\/ for the magnetic
inclination and azimuth, and 50~\ms for the LOS velocity. It is worth
noting that the deviations caused by the limited spectral sampling
are at least twice as large as those associated with instrumental
degradation effects (dotted lines in Fig.~\ref{invfull55:cap9}; see
also the middel panels of Fig.~\ref{invfull:cap9}). 

In summary, the results show that both lines are suitable for magnetic
field and velocity determinations based on ME inversions, although the
deviations may be large for the field inclination and azimuth due
to the weakness of the linear polarization signals.  Among the two,
however, the line at 525.02~nm is to be preferred because of its
larger signals and smaller deviations. 

\subsection{Uncertainties of the IMaX spectral line}
 \label{sect3.2}

  \begin{figure}[!t] \centering
   \resizebox{\hsize}{!}{\includegraphics[bb= -10 0 494 360]{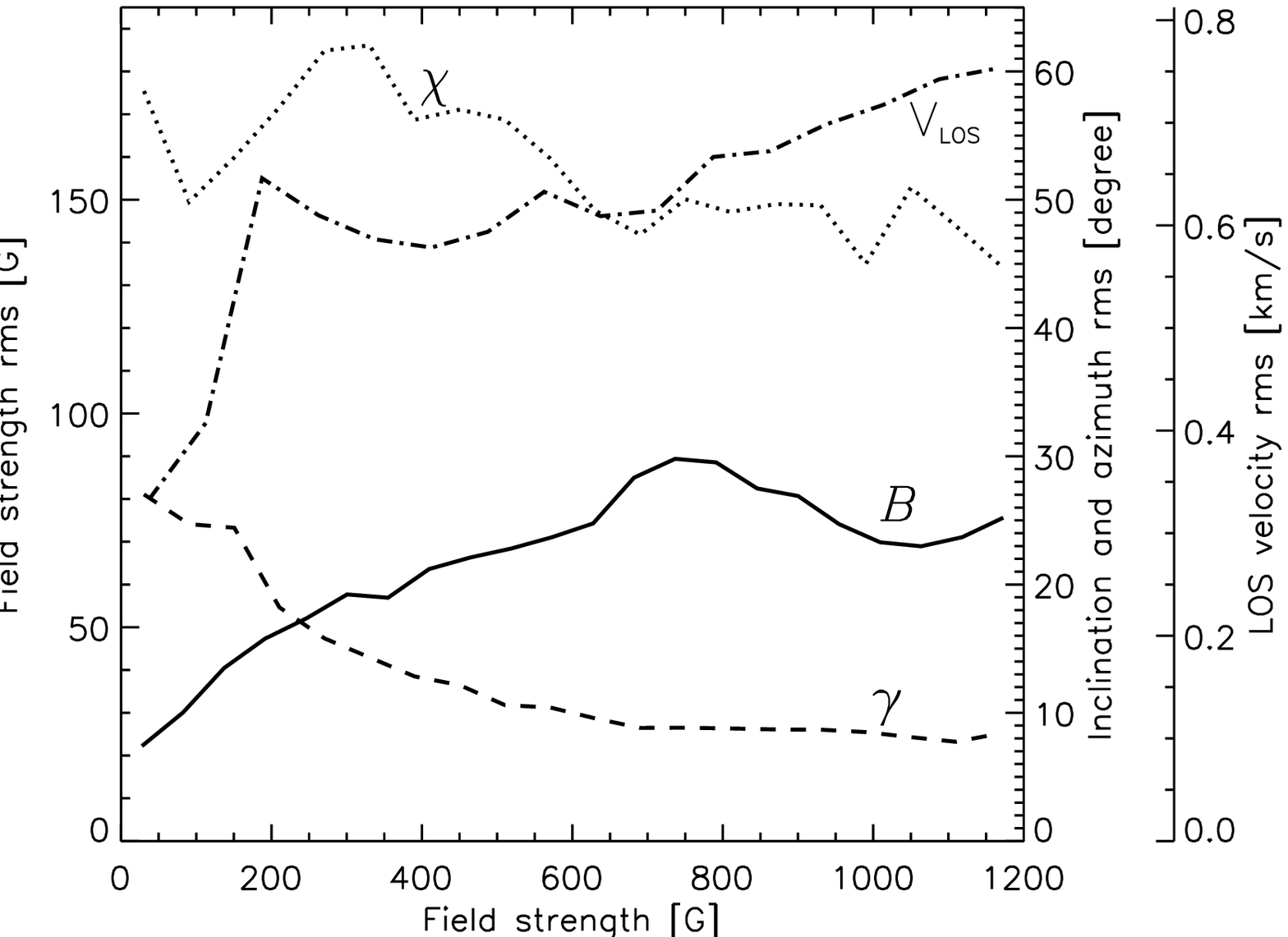}}
   \caption{Rms differences between the ME parameters inferred from
     the inversion of the \ion{Fe}{i} line at 525.02~nm as observed by
     IMaX and the corresponding MHD parameters at an optical depth of
     $\log\tau=-1.5$. The different line shapes stand for different ME
     parameters: solid - magnetic field strength, dashed - field
     inclination, dotted - field azimuth, and dot-dashed - LOS
     velocity.}
\label{rmsrms} \end{figure}  

The total uncertainties expected for the ME parameters inferred from
the IMaX vector spectropolarimetric mode with the \ion{Fe}{i} line at
525.02~nm can be calculated directly by comparing the inversion
results with the MHD models. Figure \ref{rmsrms} shows the standard
deviation of the differences between the ME parameters and the
corresponding MHD values at an optical depth\footnote{We have chosen
  the optical depth leading to the smaller rms differences.} of
$\log\tau=-1.5$.  Since an IMaX pixel is equivalent to four pixels in
the MHD simulations, the latter have been binned by a factor of two in
both directions to allow comparisons. For the magnetic field strength,
the rms values are never larger than 100~G. For the field inclination,
the uncertainty increases as the field weakens, reaching a maximum of
$\sim$~30\degree. The rms values for the field azimuth are larger, on
the order of 30-40\degree. In the case of the LOS velocity, the rms 
decreases for weak fields from 0.6 down to 0.6~\kms.

There are three sources of error contributing to the values shown in
Fig.~\ref{rmsrms}: the uncertainties caused by the instrumental
trade-offs (spectral smearing, wavelength sampling, and noise) as
given in Fig.~\ref{invfull55:cap9}, the uncertainties associated with
the spatial degradation, and the limitations of the ME approximation.
The rms differences displayed in Fig.~\ref{rmsrms} are largely
dominated by the limitations of the ME analysis.

A detailed study of the uncertainties associated with ME inversions
can be found in \cite{2010arXiv1005.5012O}. These authors analyzed
noise-free synthetic observations computed from the same MHD models we
use here, assuming very high spectral resolution and supercritical
wavelength sampling\footnote{The study was carried out for the pair of
  \ion{Fe}{i} lines at 630~nm.}. They concluded that ME inversions
provide atmospheric parameters that are averages of the actual
parameters along the line of sight. Their results show that the
uncertainties due to the ME approximation are larger than those
associated with photon noise.

\section{Summary and conclusions}
\label{sec:conclu}

We have presented an analysis of the reliability of Milne-Eddington
inversions applied to high resolution magnetograph observations such
as those delivered by IMaX. The analysis was indeed meant to help
during the design and development phases of the instrument.  In
particular, the results of this work enabled the IMaX team to select
the \ion{Fe}{i} line at 525.02 nm instead of \ion{Fe}{i} 525.06 nm.
Both lines could be used through the IMaX pre-filter, but the former
permits more accurate inferences of the vector magnetic field.
\ion{Fe}{i} 525.02 is also advantageous for the detection of the
weakest linear polarization signals of the quiet Sun.

Our analysis relies on synthetic Stokes spectra emerging from
realistic MHD simulations of the quiet Sun. The spectra were degraded
by taking into account all the {\sc Sunrise}/IMaX instrumental
effects, including telescope diffraction and CCD pixelation, smearing
by the spectral PSF of the Fabry-P\'erot \'etalon, and coarse spectral
sampling (the IMaX vector spectropolarimetric mode scans the
polarization profiles at just four wavelengths within the line and one
continuum point). Finally, the simulated IMaX observations were
inverted in terms of Milne-Eddington atmospheres.

Our main conclusion is that ME inversions can provide reasonably
accurate atmospheric parameters (especially the magnetic field
strength and the LOS velocity) from the resolved but coarsely sampled
Stokes spectra acquired by {\sc Sunrise}/IMaX.  This confirms the
conclusions of previous works \citep[e.g.,][]{2002SoPh..208..211G}.
As a by-product, we have demonstrated the important effects of pure
telescope diffraction on spectral profiles coming from the highly
inhomogeneous, quiet-Sun photosphere.


\acknowledgements We thank A. V\"ogler for kindly providing us 
with his MHD simulations. This work has been partially funded by the
Spanish Ministerio de Educaci\'on y Ciencia, through projects
ESP2006-13030-C06-02, ESP2006-13030-C06-01, AYA2009-14105-C06-06,
AYA2009-14105-C06-03, and by Junta de Andaluc\'{\i}a through project
P07-TEP-2687, including a percentage from European FEDER funds.

\bibliographystyle{aa}

\end{document}